\documentclass{INTERSPEECH2023}

\interspeechcameraready

\usepackage{amsmath,graphicx}
\usepackage{hyperref}
\usepackage{url}
\usepackage[utf8]{inputenc} 
\usepackage[T1]{fontenc}    
\usepackage{booktabs}       
\usepackage{amsfonts}       
\usepackage{nicefrac}       
\usepackage{microtype}      
\usepackage{xcolor}         
\usepackage{amsmath}
\usepackage{amssymb}
\usepackage{mathtools}
\usepackage{makecell}
\usepackage{caption}
\usepackage{stackrel}
\usepackage{enumitem}
\usepackage{multirow}

\definecolor{Red}{rgb}{0.768, 0.054, 0.054}
\definecolor{Blue}{rgb}{0.152, 0.294, 0.925}
\definecolor{Green}{rgb}{0,0.4,0.7}
\hypersetup{
    colorlinks=true,
    citecolor=teal,
    linkcolor=Red,
    urlcolor=Green,
}
\definecolor{red}{RGB}{205,33,42}

\usepackage{amsmath,amsfonts,bm}

\def\eqref#1{equation~\ref{#1}}

\def\1{\bm{1}}

\def\ve{{\bm{e}}}

\def\vs{{\bm{s}}}

\def\vx{{\bm{x}}}

\def\mI{{\bm{I}}}

\def\mY{{\bm{Y}}}

\DeclareMathAlphabet{\mathsfit}{\encodingdefault}{\sfdefault}{m}{sl}
\SetMathAlphabet{\mathsfit}{bold}{\encodingdefault}{\sfdefault}{bx}{n}

\newcommand{\bs}{\boldsymbol}

\title{ZET-Speech: Zero-shot adaptive Emotion-controllable \\ Text-to-Speech Synthesis with Diffusion and Style-based Models}
\name{Minki Kang$^{1,2*}\thanks{*: Equal Contribution}$, Wooseok Han$^{1*}$, Sung Ju Hwang$^{2}$, Eunho Yang$^{1,2}$}
\address{AITRICS$^1$, KAIST$^2$}
\small\email{\{zzxc1133,hwrg\}@aitrics.com, \quad \{sjhwang82,eunhoy\}@kaist.ac.kr}

\begin{document}

\maketitle
 
\begin{abstract}
Emotional Text-To-Speech (TTS) is an important task in the development of systems (e.g., human-like dialogue agents) that require natural and emotional speech.
Existing approaches, however, only aim to produce emotional TTS for seen speakers during training, without consideration of the generalization to unseen speakers.
In this paper, we propose ZET-Speech, a zero-shot adaptive emotion-controllable TTS model that allows users to synthesize any speaker's emotional speech using only a short, neutral speech segment and the target emotion label.
Specifically, to enable a zero-shot adaptive TTS model to synthesize emotional speech, we propose domain adversarial learning and guidance methods on the diffusion model.
Experimental results demonstrate that ZET-Speech successfully synthesizes natural and emotional speech with the desired emotion for both seen and unseen speakers.
Samples are at \url{https://ZET-Speech.github.io/ZET-Speech-Demo/}.
\end{abstract}

\section{Introduction}

\begin{figure}
    \centering
    \includegraphics[width=1.0\linewidth]{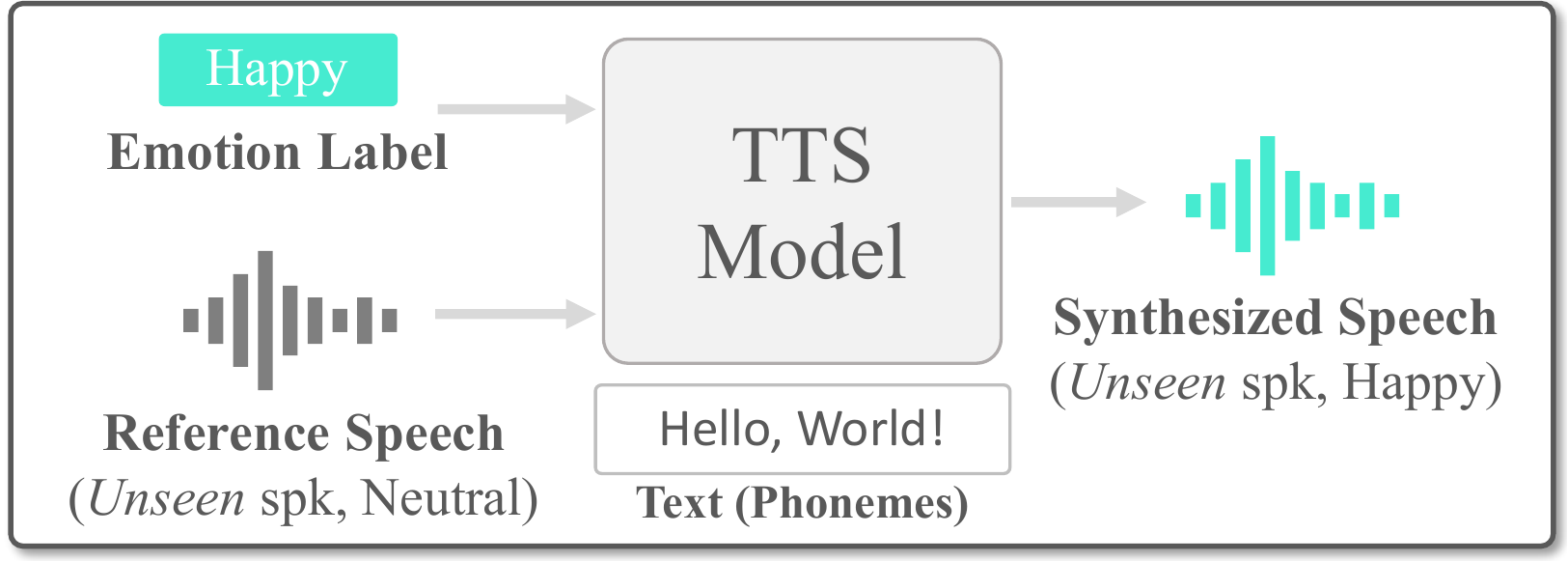}
    \vspace{-0.22in}
    \caption{\textbf{Concept.} 
    Our objective is to build a zero-shot adaptive TTS model capable of synthesizing emotional speech for both seen and unseen speakers based on emotion labels and neutral reference speech of the target speaker.}
    \vspace{-0.15in}
    \label{fig:concept}
\end{figure}

Modern Text-To-Speech (TTS) systems can now synthesize high-quality speech of a single or multiple speakers~\cite{VITS, GradTTS, VALLE}.
Beyond synthesizing the speech of seen speakers during training, several works have proposed zero-shot adaptive TTS models based on style-based generators~\cite{NeuralVoiceCloning, AdaSpeech4, StyleSpeech} and diffusion models~\cite{Grad-StyleSpeech, Guided-TTS-2}, which can generalize to unseen speakers' voice as well. 
However, the existing methods in adaptive TTS face significant limitations when applied to emotional speech generation, whose objective is to synthesize the speech with the desired emotion, in zero-shot scenarios. 
Specifically, existing approaches for zero-shot adaptive TTS do not consider the emotion-controllable generation, while the most of existing approaches for emotional TTS~\cite{StyleToken,cs2,cs3,cs4,eic1,eic2} do not take zero-shot scenarios into account. 
These limitations necessitate the development of new methods that can effectively address both requirements in TTS systems, as depicted in Figure~\ref{fig:concept}. 

In this work, we tackle the aforementioned limitations in building zero-shot adaptive emotion-controllable TTS systems. 
We assume that the emotional speeches of multiple speakers with the emotion label are given in the training stage, but only the neutral speech of the target speaker is available in the inference stage. 
In the context of training, we empirically find that a naive application of the existing emotion control methods~\cite{StyleToken, emotioncontrol} is ineffective since the emotion feature and speaker identity are highly entangled in the style vector used by the style-based generator (See Figure~\ref{fig:style-vector} for qualitative analysis). 
In order to overcome this limitation, we use domain adversarial training~\cite{DAT} to disentangle the emotional content from the style vector and make the style-based generator solely pay attention to the specified emotion condition.
We further suggest using the guidance methods~\cite{SDE, CFG} on the diffusion model in the inference stage, to improve the model's ability to synthesize the emotional speech of the target speaker conditioned on the emotion label.

Combining the above ideas, we propose a \textbf{Z}ero-shot adaptive \textbf{E}motion-controllable \textbf{T}TS model, \textbf{ZET-Speech}, that can synthesize the emotional speech of unseen speakers only with their neutral speech. 
In experiments, we empirically validate that our method enables the zero-shot adaptive TTS model to successfully perform the emotional TTS task.

Our contributions are summarized as follows:
\begin{itemize}[itemsep=0.5mm, parsep=1pt, leftmargin=*]
    \item We propose a zero-shot adaptive emotion-controllable TTS model named \textit{ZET-Speech} which can produce emotional speech given the neutral reference speech of the any speaker and the emotion condition.
    \item Our proposed approach involves domain adversarial training on the style vector, which enables the TTS model to be controlled by the provided emotional condition. Moreover, we leverage guidance methods on the diffusion model to enhance the level of emotional expression in the synthesized speech.
    \item Our method demonstrates favorable performance in generating emotional speech, as evidenced by results on the Korean~\cite{AIHub-emotion-seen, AIHub-emotion-unseen} and English~\cite{ESD,LibriTTS} speech dataset.
\end{itemize}
\section{Method}
\begin{figure}
    \centering
    \includegraphics[width=0.925\linewidth]{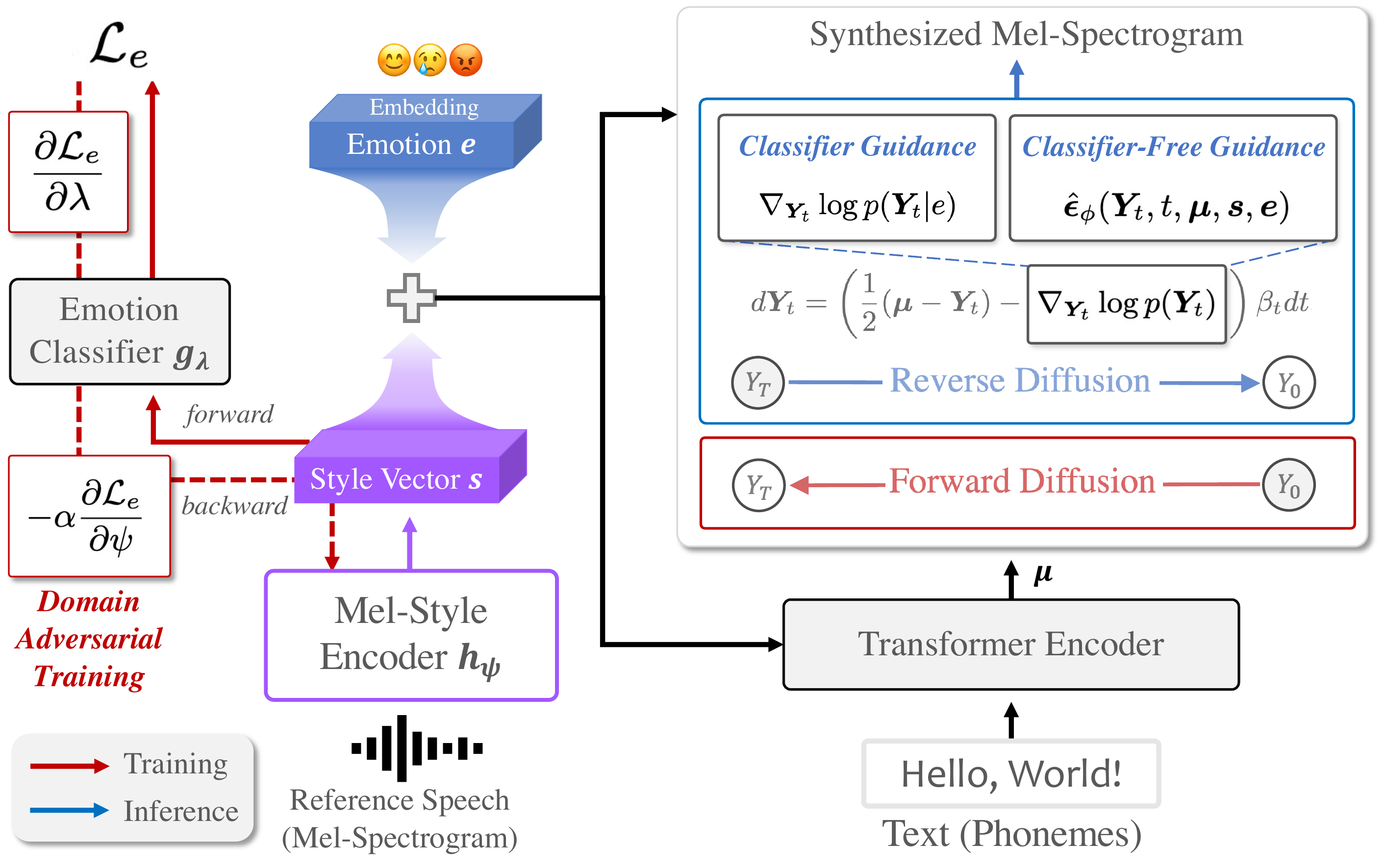}
    \caption{\textbf{Method Overview.} (Left) During training, we use domain adversarial training to separate emotion from style vectors. 
    (Right) In inference, we use guidance methods on the diffusion model to generate speech with more emotional features.}
    \vspace{-0.15in}
    \label{fig:method}
\end{figure}

As illustrated in Figure~\ref{fig:concept}, the zero-shot adaptive emotion controllable Text-to-Speech (TTS) synthesis task aims to generate the mel-spectrograms of emotional speech $\hat{\mY}$ having the target emotion, given the following inputs: the text (phonemes) $\vx$, the reference speech $\mY$, and the emotion label $e$. 
In this work, we focus on TTS models with the style-based generator~\cite{StyleGAN, StyleSpeech} and the diffusion model~\cite{SDE}. 
As shown in Figure~\ref{fig:method}, our ZET-Speech consists of two ideas upon the TTS models: (1) domain adversarial training on the style-based generator in training (\S~\ref{sec:dat}) and (2) guidance methods on the diffusion model in inference (\S~\ref{sec:guidance}).
We first briefly recapitulate the basic components in Section~\ref{sec:prelim}. 
Then, we introduce our proposed method in Section~\ref{sec:dat},~\ref{sec:guidance}

\subsection{Preliminary: Style-based Generator and Diffusion}
\label{sec:prelim}
A style-based generator~\cite{StyleSpeech, AdaSpeech} consists of two components: the transformer encoder $f_\theta$ and the mel-style encoder $h_{\psi}$. 
First of all, the mel-style encoder $h_{\psi}$ takes a mel-spectrogram of the reference speech $\mY$ as input and outputs the style vector $\vs = h_{\psi}(\mY)$.
The style vector $\vs$ plays an important role in zero-shot adaptive TTS by embedding the reference speech of any speaker into the latent space.
Then, the transformer model $f_\theta$ takes a phoneme sequence $\vx$ and the style vector $\vs$ as input and outputs the reconstructed mel-spectrogram $\boldsymbol{\mu} = f_\theta(\vx, \vs)$\footnote{For brevity, we regard the variance adaptor for the pitch, energy, and duration prediction as a part of the transformer encoder.}.

Moreover, some previous works~\cite{GradTTS, Grad-StyleSpeech} propose to generate high-fidelity speech with the diffusion models based on $\boldsymbol{\mu}$ from the generator. 
Specifically, given $\boldsymbol{\mu}$, the reverse diffusion process generates a speech by denoising the noise drawn from the gaussian distribution $\mY_T \sim \mathcal{N}(\boldsymbol{\mu}, \mI)$.
For the reverse diffusion, we define the differential equation as follows~\cite{ GradTTS,SDE}:
\begin{equation}
\label{eqn:reverse-diffusion}
    d\mY_t = \left( \frac{1}{2}(\bs{\mu} - \mY_t) - \nabla_{\mY_t} \log p(\mY_t) \right) \beta_t dt,
\end{equation}
where the neural network $ \bs{\epsilon}_\phi(\mY_t, t, \bs{\mu}, \vs)$ estimates the score function of the noisy data distribution $\nabla_{\mY_t} \log p_t(\mY_t)$ given timestep $t$, the noisy data $\mY_t$ at $t$, noise prior $\bs{\mu}$, and the style vector $\vs$~\cite{Grad-StyleSpeech}. 
For emotion-controllable TTS, we additionally input \textit{emotion vector} $\ve$ into both the hierarchical transformer encoder $f_\theta(\vx, \vs, \ve)$ and the score estimator $\bs{\epsilon}_\phi(\mY_t, t, \bs{\mu}, \vs, \ve)$.

\subsection{Domain Adversarial Training}
\label{sec:dat}

The conventional approach for incorporating emotional information into the TTS model is to concatenate or add the emotion vector to the style vector, and then forward the combined vector to the style-based generator~\cite{emotioncontrol}. 
However, this approach has proven to be ineffective in style-based generators, primarily due to the fact that the style vector $\vs$ contains both the speaker identity and emotional features of the reference speech. 
As a result, the naive method of combining the emotion vector with the style vector fails to isolate the emotional features of the reference speech and also fails to incorporate the desired emotion condition, leading to suboptimal emotional TTS performance.
For instance, as depicted in Figure~\ref{fig:concept}, assume the case where the neutral reference speech of the unseen speaker is given with the desired emotion label `happy'.
If the emotional feature is entangled in the style vector, the TTS model generates neutral speech conditioned on the neutral emotion inside the reference speech, ignoring the given emotion condition `happy'. 
Hence, it is of utmost importance to disentangle the speaker identity and emotional features from the style vector, so that the style vector solely encompasses speaker-related information.

To this end, we use domain adversarial training~\cite{DAT} to make the mel-style encoder separate the emotional features of the reference speech from the style vector. 
Specifically, we introduce an emotion classifier $g_\lambda$ which is trained to predict the emotion label of the speech given the style vector $p(e|\vs;\lambda)$, but we insert the gradient reversal layer prior to the emotion classifier which scales the gradients by the hyperparameter $-\alpha$.
Therefore, given the emotion classification loss function $\mathcal{L}_{e} = -\log p(\hat{e}|\vs;\lambda)$ with the emotion label $\hat{e}$, the gradient for the mel-style encoder parameter $\psi$ with regards to the emotion classification loss $\mathcal{L}_{e}$ becomes $- \alpha \frac{\partial \mathcal{L}_{e}}{\partial \psi}$. 
By employing DAT, the mel-style encoder embeds the speech into the style vector, which contains much less emotional information than it would have otherwise.

\subsection{Guidance Methods on the Diffusion Model}
\label{sec:guidance}
Thanks to domain adversarial training on the style vector, we can synthesize emotional speech conditioned on the provided emotion label.
The remaining issue is to make the TTS model produce more emotional speech conditioned on the given emotion.
Recently, diffusion models have demonstrated state-of-the-art performance in conditional generation~\cite{LDM}.
The guidance methods -- classifier~\cite{guided-diffusion} and classifier-free guidance~\cite{CFG} -- which allow the diffusion model to generate better-conditioned data are at the heart of a powerful conditional generation performance.

We incorporate guidance methods so that the diffusion model generates more emotional speech that is controlled by the given condition.
For classifier guidance, we train the emotion-unconditional score estimator $\bs{\epsilon}_\phi(\mY_t, t, \bs{\mu}, \vs)$ and an emotion classifier on the noisy mel-spectrograms $p(e|\mY_t)$ afterwards~\cite{EmoDiff}.
Then, we add the classifier gradient into the sampling process:
\begin{gather}
    \label{eqn:cg}
    d\mY_t = \left( \frac{1}{2}(\bs{\mu} - \mY_t) - \nabla_{\mY_t} \log p(\mY_t | e) \right) \beta_t dt, \\
    \nabla_{\mY_t} \log p(\mY_t | e) = \gamma * \nabla_{\mY_t} \log p(e | \mY_t) + \nabla_{\mY_t} \log p(\mY_t), \nonumber 
\end{gather}
where $e$ is the emotion label 
and $\gamma$ is a hyperparameter for guidance control.

\begin{table*}

	\centering
        \setlength{\tabcolsep}{1.5em}
	\small
        \caption{\textbf{Objective evaluation on Seen speakers (Korean).} Experimental results on speakers where the models are trained on. Note that the classifier guidance is not applicable along with the Global Style Token.}
        \vspace{-0.1in}
	\resizebox{1.0\textwidth}{!}{
	\begin{tabular}{lccc|ccc}
		\toprule
  & \multicolumn{3}{c}{\textbf{+ Emotion Label}} & \multicolumn{3}{c}{\textbf{+ Global Style Token (GST)}}  \\
		{\textbf{Model}} & ECA ($\uparrow$)       & CER ($\downarrow$) &  SECS ($\uparrow$) & ECA ($\uparrow$)    & CER ($\downarrow$)  &  SECS ($\uparrow$)  \\
		\midrule[0.8pt]  
		\textbf{Grad-StyleSpeech} & 16.35$_{\pm \text{8.40}}$ & 16.84$_{\pm \text{11.28}}$ & \bf 0.778 & 12.54$_{\pm \text{8.20}}$ & 17.60$_{\pm \text{11.65}}$  & \bf 0.778  \\
		\textit{+ Domain Adversarial Training (DAT)} & 31.27$_{\pm \text{15.78}}$ & \bf 11.93$_{\pm \text{9.48}}$ & 0.757 & 24.44$_{\pm \text{16.94}}$ & \bf 10.99$_{\pm \text{9.06}}$ & 0.746   \\
            \textit{+ Classifier Guidance (CG)} & 28.73$_{\pm \text{14.04}}$ & 21.51$_{\pm \text{20.08}}$ & 0.765 & - & - & -  \\
		\textit{+ Classifier-free Guidance (CFG$_{\gamma=1.75}$)} & 19.68$_{\pm \text{10.12}}$ & 26.38$_{\pm \text{18.15}}$ & 0.776 & 13.81$_{\pm \text{9.15}}$ & 23.21$_{\pm \text{16.33}}$ & 0.776  \\
            \midrule
            \textit{Ours} \\
            \textbf{ZET-Speech} \textit{(CG)} & \bf 51.59$_{\pm \text{16.45}}$  & 16.22$_{\pm \text{16.97}}$ & 0.759 & - & - & - \\
            \textbf{ZET-Speech} \textit{(CFG$_{\gamma=1.25}$)} & 41.75$_{\pm \text{14.64}}$ & 13.26$_{\pm \text{10.31}}$ & 0.750 & 33.65$_{\pm \text{15.40}}$ & 11.84$_{\pm \text{9.66}}$ & 0.770  \\
            \textbf{ZET-Speech} \textit{(CFG$_{\gamma=1.75}$)} & 49.84$_{\pm \text{17.33}}$ & 16.33$_{\pm \text{11.74}}$ & 0.747 & \bf 38.89$_{\pm \text{13.81}}$ & 15.95$_{\pm \text{14.68}}$ & 0.757  \\
		\bottomrule
	\end{tabular}
	}
 	\vspace{-0.2in}
 	\label{tab:objective-seen}
\end{table*}

\begin{table}
	\centering
        \caption{\textbf{Objective evaluation on Unseen speakers (Korean).} We use unseen speakers where the models are \textbf{not} trained on.}
        \vspace{-0.1in}
	\resizebox{1.0\linewidth}{!}{
	\begin{tabular}{lcc|cc}
		\toprule
  & \multicolumn{2}{c}{\textbf{+ Emotion Label}} & \multicolumn{2}{c}{\textbf{+ GST}}  \\
		{\textbf{Model}} & ECA ($\uparrow$)    & CER ($\downarrow$)  & ECA ($\uparrow$)     & CER ($\downarrow$)    \\
		\midrule[0.8pt]  
		\textbf{Grad-StyleSpeech} & 18.29$_{\pm \text{8.59}}$ & 10.20$_{\pm \text{9.44}}$ & 13.72$_{\pm \text{7.26}}$  & \bf 9.98$_{\pm \text{9.85}}$  \\
		\textit{+ DAT} & 26.01$_{\pm \text{12.99}}$ & \bf 9.32$_{\pm \text{9.79}}$ & 23.71$_{\pm \text{13.28}}$ & 10.15$_{\pm \text{9.35}}$   \\
            \textit{+ CG} & 23.29$_{\pm \text{13.65}}$ & 17.83$_{\pm \text{21.78}}$ & - & - \\
		\textit{+ CFG$_{\gamma=1.75}$} & 17.86$_{\pm \text{9.56}}$ & 15.74$_{\pm \text{13.67}}$ & 12.43$_{\pm \text{9.40}}$ & 15.03$_{\pm \text{14.35}}$  \\
            \midrule
            \textit{Ours} \\
            \textbf{ZET-S.} \textit{(CG)} & \bf 39.86$_{\pm \text{16.57}}$ & 12.15$_{\pm \text{15.29}}$ & - & - \\
            \textbf{ZET-S.} \textit{(CFG$_{\gamma=1.25}$)} & 31.57$_{\pm \text{14.18}}$ & 11.26$_{\pm \text{10.66}}$ & 31.29$_{\pm \text{15.61}}$ & 10.50$_{\pm \text{8.75}}$\\
            \textbf{ZET-S.} \textit{(CFG$_{\gamma=1.75}$)} & 34.57$_{\pm \text{16.56}}$ & 14.01$_{\pm \text{12.21}}$ & \bf 32.57$_{\pm \text{16.67}}$ & 13.57$_{\pm \text{13.78}}$ \\
		\bottomrule
	\end{tabular}
	}
 	\vspace{-0.15in}
 	\label{tab:objective-unseen}
\end{table}

With regard to classifier-free guidance, we randomly replace the emotion embedding with a null embedding $\varnothing$ in training. 
The important point to consider is that $\bs{\mu} = f_\theta(\vx, \vs, \ve)$ also contains emotional information. 
Therefore, we should also forward $\bs{\mu}_\varnothing$ to estimate the unconditional noise. 
We forward null embedding into a transformer encoder to generate $\bs{\mu}_\varnothing = f_\theta(\vx, \vs, \varnothing)$.
We then combine the conditional and unconditional score estimation to perform sampling with classifier-free guidance. Formally,
\begin{align}
\label{eqn:cfg}
    &\hat{\bs{\epsilon}}_\phi(\mY_t, t, \bs{\mu}, \vs, \ve) = \\
    &\bs{\epsilon}_\phi(\mY_t, t, \bs{\mu}, \vs, \ve) + \gamma (\bs{\epsilon}_\phi(\mY_t, t, \bs{\mu}, \vs, \ve) - \bs{\epsilon}_\phi(\mY_t, t, \bs{\mu}_\varnothing, \vs, \varnothing)), \nonumber
\end{align}
where $\gamma$ is a hyperparameter for guidance control.
Then, we modify the differential equation in Equation~\ref{eqn:reverse-diffusion} with score estimation in Equation~\ref{eqn:cfg} as follows:
\begin{equation}
    d\mY_t = \left( \frac{1}{2}(\bs{\mu} - \mY_t) - \hat{\bs{\epsilon}}_\phi(\mY_t, t, \bs{\mu}, \vs, \ve) \right) \beta_t dt.
\end{equation}

\section{Experiment}
\label{sec:exp}
\subsection{Experimental Setup}
\subsubsection{Implementation Details}

We use Grad-StyleSpeech~\cite{Grad-StyleSpeech} as the base model due to its promising performance based on the style-based generator and diffusion model. 
We start from the TTS model pre-trained on a neutral-only multi-speaker dataset. 
Then, we train the pre-trained TTS model on the emotional speech dataset for 200k steps with the batch size of 8 on a single TITAN RTX GPU, using an Adam optimizer with a learning rate of 1e-3 and a linear decay.

\subsubsection{Dataset}
We mainly target Korean speech synthesis since there are high-quality Korean speech datasets available from the public organization AIHub\footnote{https://aihub.or.kr/}.
We pre-train the model on the large-scale multi-speaker Korean~\cite{AIHub-multispeaker} or English~\cite{LibriTTS} dataset. 
For Korean, we train the model on the multi-emotional dataset~\cite{AIHub-emotion-seen}. 
To evaluate the unseen speaker performance, we use a separate Korean dataset~\cite{AIHub-emotion-unseen} not used in training.
For English, we use ESD~\cite{ESD} dataset for training and LibriTTS Test~\cite{LibriTTS} for evaluation.
We downsample all speech to 16kHz and use HiFi-GAN~\cite{HifiGAN} as a vocoder.

\subsubsection{Evaluation Setup}
In this subsection, we provide detailed evaluation setups.

\textbf{Baselines.}
We use Grad-StyleSpeech~\cite{Grad-StyleSpeech} as the base model for all baselines for a fair comparison.
\textbf{(1) Grad-StyleSpeech (GSS).} Vanilla Grad-StyleSpeech model where we input the emotion embedding from the hard emotion label or global style token~\cite{StyleToken} as input~\cite{emotioncontrol}.
\textbf{(2) GSS + Domain Adversarial Training.} Only applying domain adversarial training on the speaker vector.
\textbf{(3) GSS + Classifier Guidance.} Only applying classifier guidance on the diffusion model similar to EmoDiff~\cite{EmoDiff}.
\textbf{(4) GSS + Classifier-free Guidance.} Only applying classifier-free guidance on the diffusion model.
\textbf{(5) ZET-Speech (CG, ours).} Our ZET-Speech model with classifier guidance ($\gamma = 50$).
\textbf{(6) ZET-Speech (CFG, ours).} Our ZET-Speech model with classifier-free guidance ($\gamma = 1.25, 1.75$).

\textbf{Evaluation Metric.}
For objective evaluation, we introduce three metrics.
The first one is \textbf{Emotion Classifier Accuracy (ECA)} which measures how emotional the synthesized speech is. Specifically, we utilize a pre-built emotion classification model based on the Audio Spectrogram Transformer (AST)~\cite{AST} and predict the emotion of the synthesized speech.
If the AST model successfully predicts the target emotion, we regard this as a successful case of emotional TTS.
We measure \textbf{Character Error Rate (CER)} to evaluate the quality of the synthesized speech. We leverage Whisper~\cite{whisper} to recognize the text within the synthesized speech.
To check whether the synthesized speech reflects the target speaker's identity, we measure \textbf{Speaker Embedding Cosine Similarity (SECS)}.
For SECS, we measure the similarity between vectors of the synthesized and gold emotional speech.
To embed speech into the vector, we use the speaker encoder from the resemblyzer library\footnote{https://github.com/resemble-ai/Resemblyzer}.
If there is no gold emotional speech available, we cannot measure the SECS (unseen speaker case).
For subjective evaluation, we use Mean Opinion Score (MOS) to measure the emotional naturalness of the synthesized speech.
For MOS, we require the evaluator to assess the degree to which the given emotion is naturally reflected in the synthesized speech.
We also measure speaker Similarity MOS (SMOS) for seen speakers, by requesting the evaluator to check the voice similarity between the synthesized speech and the gold emotional speech.
For all experiments, we use neutral speech as a reference speech and give emotion conditions as a hard label (Emotion Label) or sampled emotional speech from the training dataset (Global Style Token).
We use 9 seen speakers and 10 unseen speakers with 10 scripts and 7 emotions for evaluations.
We recruit 20 evaluators for subjective evaluations.

\begin{table}
	\centering
	\small
        \caption{\textbf{Objective evaluations on Unseen speakers (English).} We use unseen speakers (LibriTTS Test~\cite{LibriTTS}) for evaluations.}
        \vspace{-0.1in}
	\resizebox{0.75\linewidth}{!}{
	\begin{tabular}{lcc}
		\toprule
		{\textbf{Model}}  & ECA ($\uparrow$)    & CER ($\downarrow$)        \\
		\midrule[0.8pt]  
		\textbf{GSS}     & 40.50$_{\pm\text{14.82}}$ & \bf 13.55$_{\pm\text{9.69}}$    \\
            \textbf{ZET-Speech} \textit{(CFG$_{\gamma=1.5}$, ours)} & \bf 59.75$_{\pm\text{15.65}}$ & 21.96$_{\pm\text{13.86}}$  \\
		\bottomrule
	\end{tabular}
	}
 	\vspace{-0.18in}
 	\label{tab:english}
\end{table}

\subsection{Experimental Results}
In Table~\ref{tab:objective-seen} and~\ref{tab:objective-unseen}, we present the performance of models in terms of objective evaluation metrics for seen and unseen speakers, respectively.
Results show that our ZET-Speech outperforms baselines in terms of ECA, demonstrating that ZET-Speech is able to synthesize the emotional speech conditioned on the given emotion.
These findings suggest that our method can be successfully applied to any speaker even if there is only a limited amount of neutral speech available.
To elaborate, Domain Adversarial Training (DAT) significantly enhances the model's ability to generate emotional speech. In other words, the model has lower sensitivity to emotion conditions without DAT. Furthermore, the ECA scores obtained by comparing ZET-Speech with the GSS \textit{+ DAT} baseline indicate that guidance methods are beneficial in improving the emotional expressivity of the synthesized speech.

Experimental results show that both guidance methods contribute to improving ECA scores.
Although classifier guidance shows better ECA than classifier-free guidance, classifier-free guidance has exclusive merits compared to classifier guidance where it does not require an additional classifier and can be combined with other methods like global style token~\cite{StyleToken}.

Our empirical findings also suggest that incorporating guidance methods may have a negative impact on the quality of the synthesized speech, resulting in a low CER and SECS metric. However, we also observe that there exists a trade-off between emotional expressivity and speech quality, which can be managed by adjusting the relevant hyperparameters $\gamma$. We provide an extensive analysis of this phenomenon in Section~\ref{sec:analysis}.

In Table~\ref{tab:english}, we briefly present the object evaluations on the English benchmark~\cite{ESD,LibriTTS}. 
We also upload speech samples in English on the demo page (\url{https://ZET-Speech.github.io/ZET-Speech-Demo/}).

Table~\ref{tab:subjective} shows subjective evaluation results. Most evaluators found ZET-Speech more effective in conveying desired emotoin than Grad-StyleSpeech, consistent with Tables~\ref{tab:objective-seen} and~\ref{tab:objective-unseen}. Interestingly, evaluators found the speech from ours with classifier-free guidance more natural and emotional than the speech from ours with classifier guidance, which differs from Tables~\ref{tab:objective-seen} and~\ref{tab:objective-unseen}.

\begin{table}
	\centering
	\small
        \caption{\textbf{Subjective evaluations (Korean).} Experimental results on seen and unseen speakers with human evaluation.}
        \vspace{-0.1in}
	\resizebox{0.90\linewidth}{!}{
	\begin{tabular}{lcc|c}
		\toprule
  & \multicolumn{2}{c}{\textbf{Seen}} & \multicolumn{1}{c}{\textbf{Unseen}}  \\
		{\textbf{Model}} & MOS  & SMOS  & MOS    \\
		\midrule[0.8pt]  
		\textbf{GSS}                  & 2.60$_{\pm\text{.13}}$ & 3.11$_{\pm\text{.15}}$ & 2.63$_{\pm\text{.11}}$   \\
            \textbf{ZET-Speech} \textit{(CG, ours)} & 3.02$_{\pm\text{.13}}$ & 3.12$_{\pm\text{.14}}$ & 3.06$_{\pm\text{.13}}$   \\
            \textbf{ZET-Speech} \textit{(CFG, ours)} & \bf 3.44$_{\pm\text{.14}}$ & \bf 3.22$_{\pm\text{.13}}$  & \bf 3.31$_{\pm\text{.15}}$  \\
		\bottomrule
	\end{tabular}
	}
 	\vspace{-0.2in}
 	\label{tab:subjective}
\end{table}

\begin{figure}[t]
    \centering
    \includegraphics[width=0.95\linewidth]{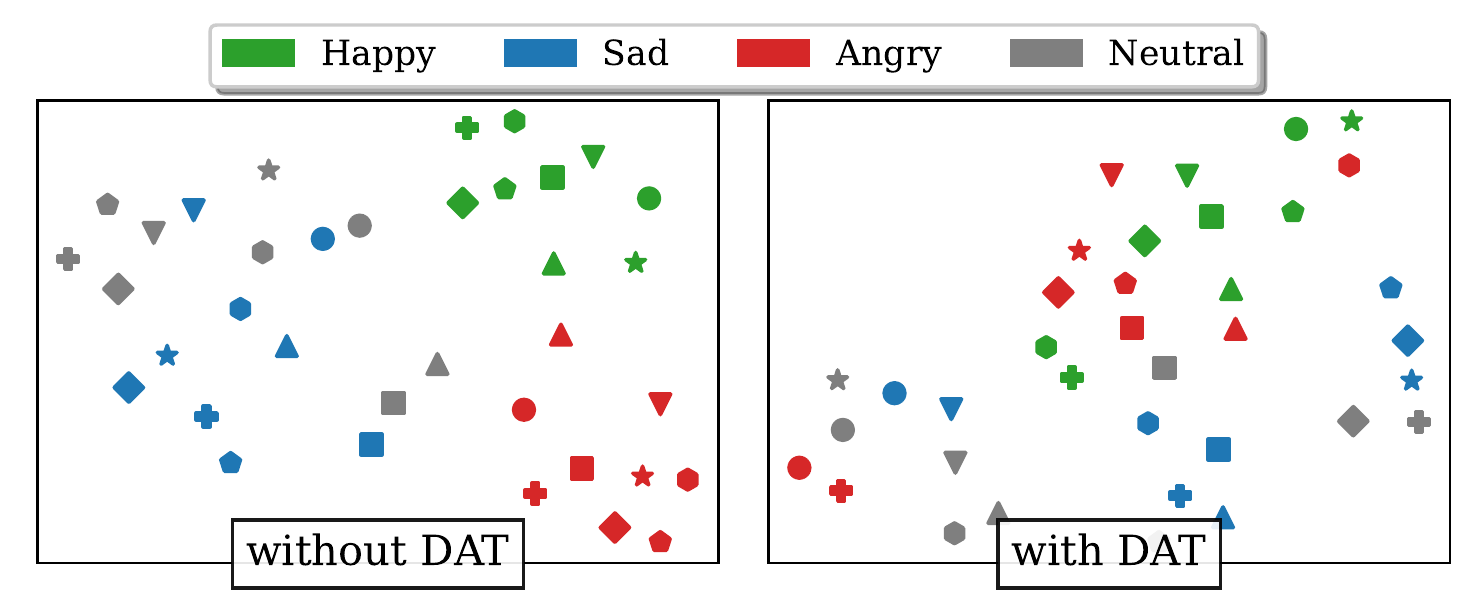}
    \vspace{-0.1in}
    \caption{\textbf{Style Vector Visualization.} To visually analyze the effect of Domain Adversarial Training (DAT), we plot style vectors from various speakers and emotions. The same marker denotes the same speaker, and the same color denotes the same emotion.}
    \vspace{-0.2in}
    \label{fig:style-vector}
\end{figure}

\subsection{Analysis}
\label{sec:analysis}

\begin{figure}[ht!]
    \centering
    \includegraphics[width=0.87\linewidth]{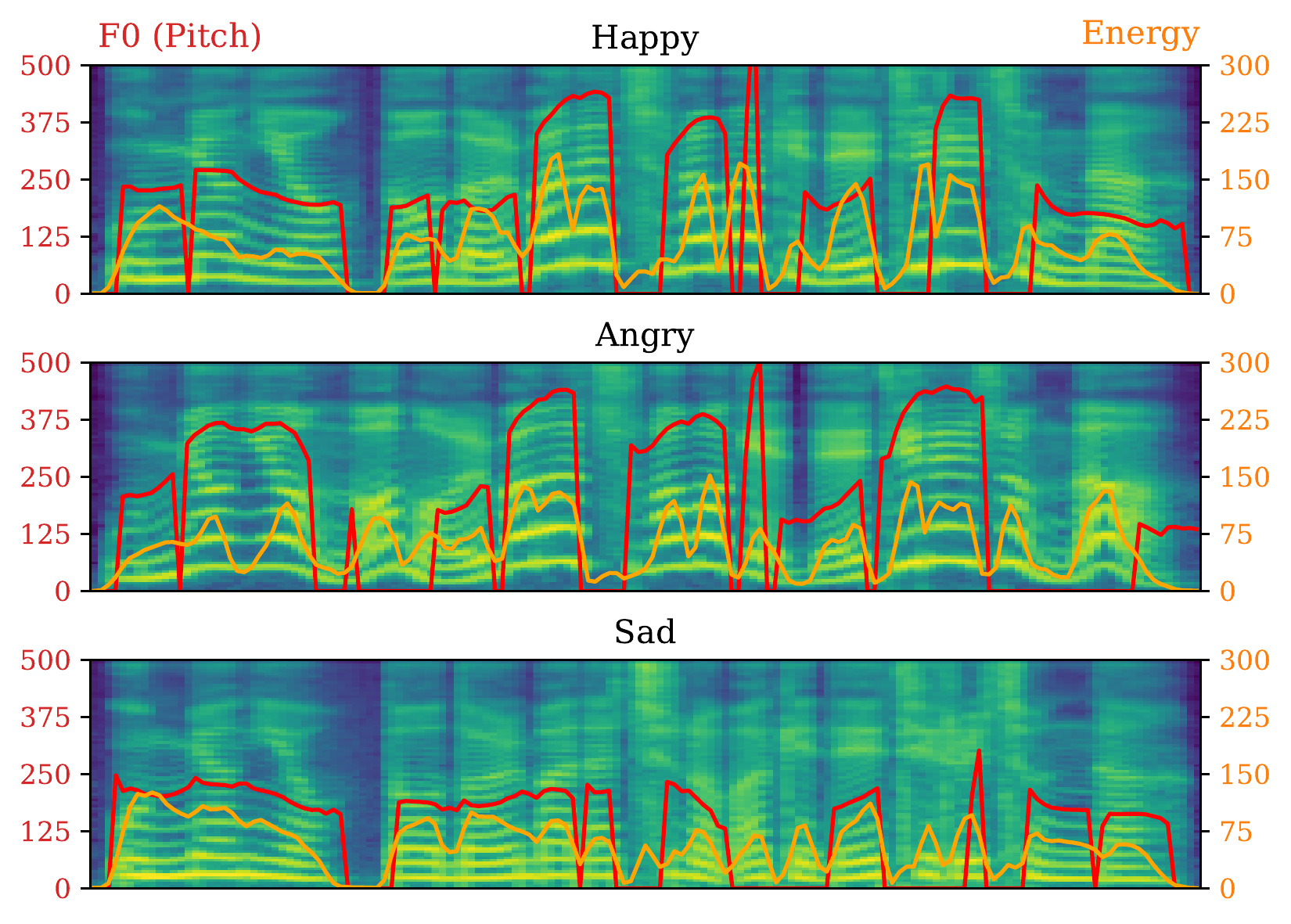}
    \vspace{-0.15in}
    \caption{\textbf{Spectrum Visualization.} We plot the mel-spectrograms, pitch, and energy contours of the synthesized speech varying the emotion condition, given the same script and unseen speaker.}
    \vspace{-0.1in}
    \label{fig:spec-plot}
\end{figure}
\begin{figure}[t]
    \centering
    \includegraphics[width=0.93\linewidth]{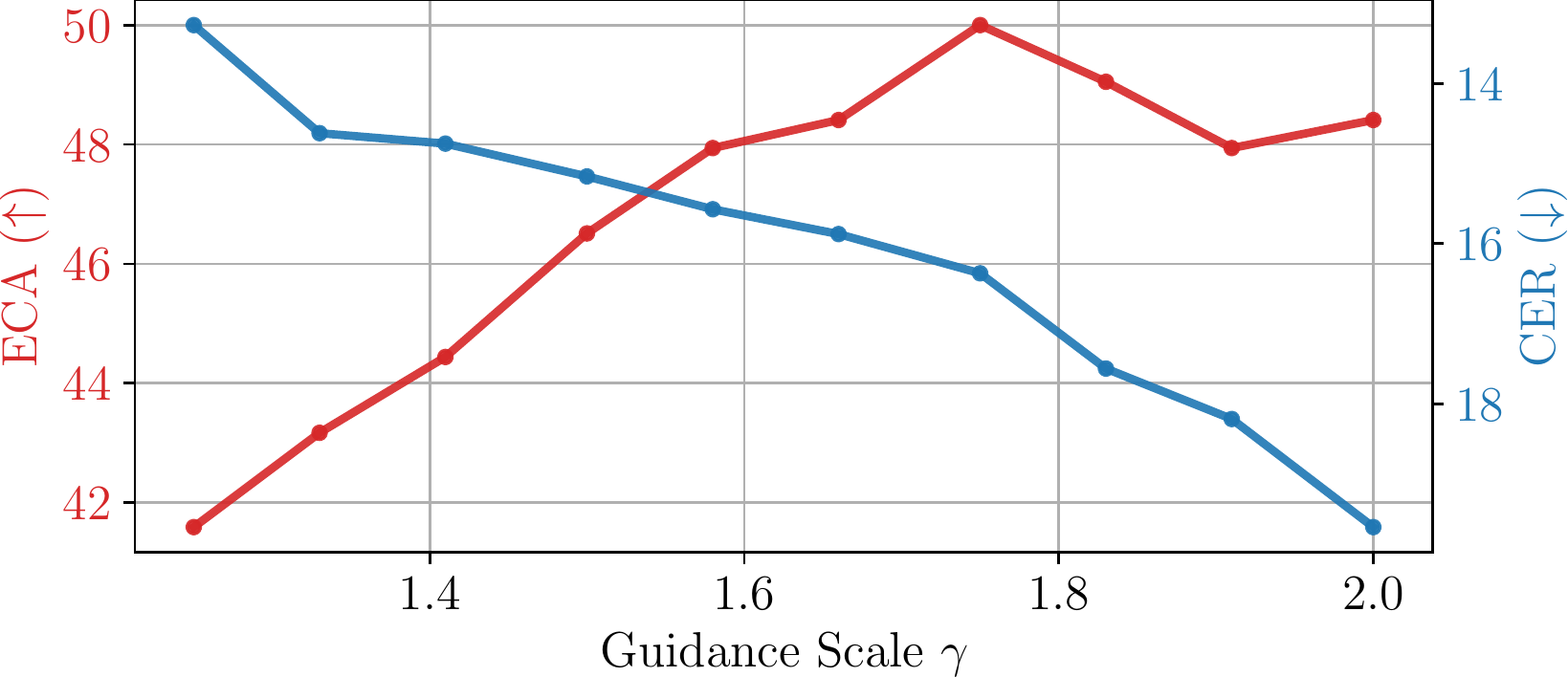}
    \vspace{-0.13in}
    \caption{\textbf{Guidance Scale Analysis.} In classifier-free guidance, we measure ECA and CER by varying the guidance scale $\gamma$.}
    \vspace{-0.22in}
    \label{fig:scale-plot}
\end{figure}

\textbf{Effect of Domain Adversarial Training.}
In Figure~\ref{fig:style-vector}, we plot the style vectors of the randomly sampled speakers from the training dataset. We sample speech with four major emotions (Happy, Sad, Angry, and Neutral) from each speaker and use t-SNE~\cite{tSNE} to reduce the dimension.
We plot the style vectors from the model trained with and without Domain Adversarial Training (DAT).
The style vectors from the model trained without DAT are clearly clustered based on emotion in Figure~\ref{fig:style-vector}, demonstrating how the emotional features are entangled with the style vector.
On the other hand, the style vectors from the model trained with DAT are more evenly distributed across the emotion and less tightly clustered. 
The results demonstrate that emotional features are successfully disentangled from the style vector with DAT.

\textbf{Spectrum Visualization.}
In Figure~\ref{fig:spec-plot}, we show the mel-spectrograms of the synthesized speech for different emotions, accompanied by pitch (F0) and energy contours to highlight the variations among them. The visualizations clearly show that the speech generated for Happy and Angry emotions typically has a higher pitch than that of Sad emotion, emphasizing the ability of our model to produce emotionally expressive speech with distinct variations for different emotional conditions. 

\textbf{Guidance Scale.}
Our analysis indicates that the guidance scale has a significant impact on both the quality and emotional expressiveness of synthesized speech in guidance methods. To evaluate this trade-off, we measure the Emotional Classification Accuracy (ECA) and Character Error Rate (CER) while varying the guidance scale $\gamma$ from $1.25$ to $2.0$ in ZET-Speech with classifier-free guidance.
As illustrated in Figure~\ref{fig:scale-plot}, we observe that increasing the guidance scale results in higher ECA and CER. This implies that a higher guidance scale improves emotion expressivity, but at the expense of reduced speech quality.

\section{Conclusion}
We proposed ZET-Speech, a zero-shot adaptive emotion controllable TTS model capable of generating emotional speech in any speaker's voice only with neutral reference speech and emotion labels. 
Our approach employs domain adversarial training to disentangle emotion features from the style vector during training, allowing the synthesized speech to be controlled solely by the emotion condition. Furthermore, we utilize guidance methods on the diffusion model to enhance the control of emotion in a synthesized speech.
Experimental results demonstrated that our method significantly outperforms baselines in objective and subjective measures of emotional expressivity. We hope that our method enables practitioners to synthesize emotional speech for desired speakers without the need for emotional recordings.





\bibliographystyle{IEEEtran}
\bibliography{mybib}

\end{document}